\documentclass{article}
\usepackage{spconf,amsmath,graphicx}

\usepackage{url}
\usepackage{color}
\usepackage[english]{babel}
\usepackage[T1]{fontenc}
\usepackage[utf8]{inputenc}
\usepackage{booktabs} 
\usepackage{algorithm}
\usepackage{algorithmic}

\title{Contrastive Unsupervised Learning for Audio Fingerprinting}

\name{Zhesong Yu, Xingjian Du, Bilei Zhu, Zejun Ma}
\address{Bytedance AI Lab}

\begin{document}
%
\maketitle
\begin{abstract}
The rise of video-sharing platforms has attracted more and more people to shoot videos and upload them to the Internet. These videos mostly contain a carefully-edited background audio track, where serious speech change, pitch shifting and various types of audio effects may involve, and existing audio identification systems may fail to recognize the audio. To solve this problem, in this paper, we introduce the idea of contrastive learning to the task of audio fingerprinting (AFP). Contrastive learning is an unsupervised approach to learn representations that can effectively group similar samples and discriminate dissimilar ones. In our work, we consider an audio track and its differently distorted versions as similar while considering different audio tracks as dissimilar. Based on the momentum contrast (MoCo) framework, we devise a contrastive learning method for AFP, which can generate fingerprints that are both discriminative and robust. A set of experiments showed that our AFP method is effective for audio identification, with robustness to serious audio distortions, including the challenging speed change and pitch shifting.

\end{abstract}
\begin{keywords}
Audio fingerprinting, audio identification, unsupervised learning, contrastive learning.
\end{keywords}
\section{Introduction}
\label{sec:intro}
Audio fingerprinting (AFP) is at the core of content-based audio identification. Given a short and unlabeled audio snippet, an AFP system first extracts a compact signature, i.e., a fingerprint, to summarize the audio, and then matches the fingerprint with those calculated in advance from a set of reference audio. If a match is found, the query is identified. In the last two decades, various AFP algorithms have been proposed. Examples include the Philips robust hash (PRH) \cite{haitsma2002highly}, the landmark-based method \cite{wang2003industrial}, the Waveprint system that employs wavelet for AFP \cite{baluja2007audio} and the masked audio spectral keypoints (MASK) method \cite{anguera2012mask}. Also, there exists a number of commercial applications of AFP, including Shazam\footnote{\url{https://www.shazam.com}}, SoundHound\footnote{\url{https://www.soundhound.com}} and Musicxmatch\footnote{\url{https://www.musixmatch.com}}, among others.

In the real-world environment, the audio to be identified may be subject to various types of degradation during transmissions, such as compression, background noise, echo, equalization and speech change. To allow an accurate identification of these audio, AFP systems should be robust to real-world environmental distortion and interference \cite{cano2005review}. Moreover, in recent years, with the popularity of video-sharing platforms such as Youtube\footnote{\url{https://www.youtube.com/}} and Tiktok\footnote{\url{https://www.tiktok.com}}, more and more videos are generated and uploaded everyday. When producing a video, it is common that people choose an audio track and carefully edit it to fit the video and serve as the background. The artificial editing of these audio tracks may include serious time scaling, pitch shifting, and adding a variety of audio effects, making AFP-based audio identification more challenging.

Recently, there is a new trend in AFP research that deep learning, especially unsupervised deep learning, has been introduced for audio identification. In \cite{gfeller2017now}, y Arcas \textit{et. al} described Google's implementation of a low-power music recognizer, which extracts fingerprints using a stack of convolutional layers, followed by a two-level divide-and-encode block. In \cite{baez2020samaf}, B{\'a}ez-Su{\'a}rez \textit{et. al} proposed SAMAF, which uses a sequence-to-sequence autoencoder model consisting of long-short term memory (LSTM) layers to generate audio fingerprints. Experiments in \cite{baez2020samaf} showed the advantage of SAMAF over traditional AFP methods such as \cite{haitsma2002highly} and \cite{wang2003industrial} in most cases. However, the robustness of SAMAF to certain audio distortions such as pitch shifting and speed change is still weak, limiting its use for audio in video-sharing platforms.

In this work, we follow the idea of using unsupervised learning for AFP. Specifically, we introduce the emerging \emph{contrastive learning} \cite{hadsell2006dimensionality} to the AFP task to solve the audio identification problem. Contrastive learning is a technique that aims at learning representations that can effectively group similar samples and discriminate dissimilar ones, typically in an unsupervised manner \cite{hadsell2006dimensionality}. To achieve this goal, a number of contrastive learning methods have been proposed \cite{wu2018unsupervised,ye2019unsupervised,oord2018representation,hjelm2018learning,bachman2019learning}, and examples include the momentum contrast (MoCo) methods \cite{he2020momentum, chen2020improved} and the simple framework of contrastive learning of visual representations (SimCLR) \cite{chen2020simple}. These methods have achieved promising results in several computer vision tasks such as object detection and image segmentation, with even superior performance than their supervised pre-training counterparts in some cases.

In this paper, we propose an AFP system based on MoCo \cite{he2020momentum}, which is one of the state-of-the-art methods of contrastive learning. In our use of MoCo, each audio track and its differently distorted versions are considered similar, while different audio tracks are deemed dissimilar. During training, a contrastive loss is optimized when the distances between the representations of similar audio are minimized and those between dissimilar audio are maximized. In this case, representations that are both robust and discriminative can be learned and these representations are used as fingerprints for audio identification. Please note that since our model employs contrastive learning, the robustness of our model to any audio degradation can be learned (including serious pitch shifting and speech change for which the SAMAF method is less robust) by simply applying the target degradation to each audio and adding the distorted audio into the training set.

\section{Approach}
\label{sec:approach}
In this section, we first describe our contrastive unsupervised learning method of training a fingerprint encoder. Then, we discuss the details of using the encoder to extract fingerprints from audio, and the details of fingerprint matching for audio identification. 

\subsection{Contrastive Learning for Audio Fingerprinting}
\label{sec:contrast_learning}

Contrastive learning is an unsupervised (or ``self-supervised'' in some works) approach to representation learning, which attempts to learn useful data embeddings by training on similar and dissimilar pairs of data. The similar and dissimilar pairs of data are generated by maintaining a \emph{dynamic dictionary} and given an encoded query $q$, the dictionary maintains a single key $k^+$ that is similar to $q$ and a set of keys $\{k^-\}$ that are dissimilar to $q$. Hereafter, we call $k^+$ a positive sample and $\{k^-\}$ negative samples\footnote{In this paper, we use the form $\{x\}$ to represent a set of data samples, where $x$ is a single sample.}.

\begin{figure}[t]
    \centering
    \includegraphics[trim = 1mm 1mm 1mm 1mm , clip, width=2.4in]{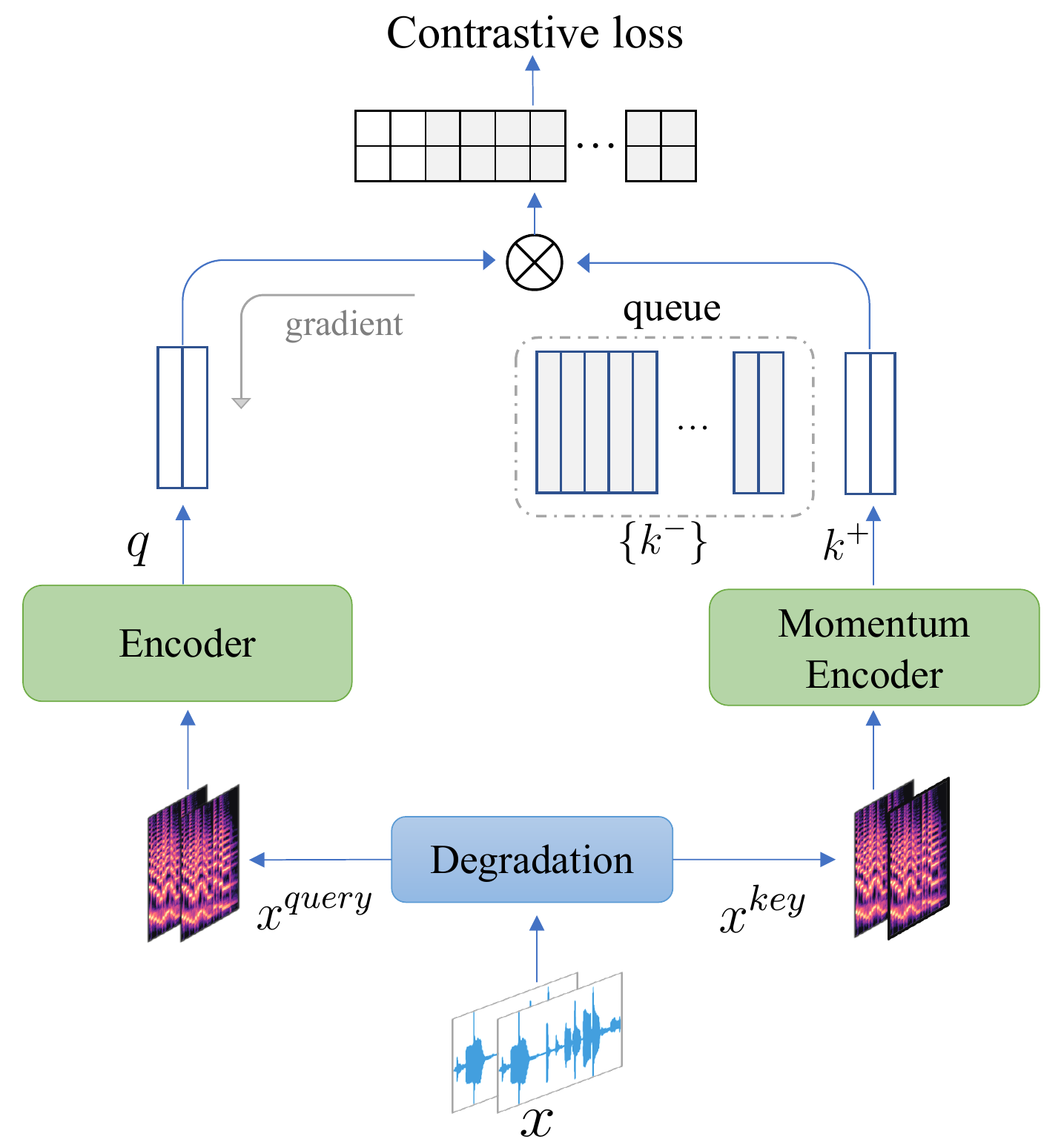}
    \caption{The training mechanism of MoCo-based AFP.}
    \label{fig:moco}
\end{figure}

In our use of contrastive learning for AFP, an audio track and its differently distorted versions are considered similar (being positive samples), while different audio tracks are deemed dissimilar (being negative samples). Fig. \ref{fig:moco} shows the mechanism of our contrastive learning. First, a batch of $N$ audio recordings is given as input. We perform random audio degradation for each audio in the batch and cut a snippet of length $T$ from the beginning of the degraded audio. These fixed-length audio snippets form a new audio batch $\{x^{query}\}$, and we repeat the above operation to generate a third audio batch $\{x^{key}\}$. Obviously, for each audio $x^{query}$, there exists a single positive sample in $\{x^{key}\}$, as well as $N - 1$ negative samples. Then, each audio $x^{query}$ in $\{x^{query}\}$ is transformed into a Mel-spectrogram and goes through an encoder to obtain an embedding $q$. The audio snippets $\{x^{key}\}$ are also encoded into a set of embeddings $\{k\}$. Please note that similar to the audio case, for each $q$, there is a single positive sample and $N-1$ negative samples in $\{k\}$. Afterward, the embedding set $\{k\}$ is integrated into a dictionary as new keys.

The size of the dictionary needs to be large (significantly larger than a common batch size) to ensure a better sampling of the underlying data space. This poses a challenge to the maintenance of the dictionary. To solve this problem, we employ MoCo \cite{he2020momentum} and consider the dictionary as a queue of data samples. When a new batch of $N$ keys come, we enqueue the batch to the dictionary and remove the oldest $N$ keys from the queue. By this means, the dictionary is progressively updated and the size of the dictionary remains fixed. Also, adding $\{k\}$ into the queue ensures that there is a positive sample $k^+$ in the dictionary for each $q$ in the current batch. All the other keys $\{k^-\}$, including the $N-1$ keys in the current batch excluding $k^+$, as well as all the past keys in the dictionary, are negative samples.

The goal of our contrastive learning is to optimize a contrastive loss, i.e., the InfoNCE \cite{oord2018representation}, which has been widely used in related studies. The InfoNCE loss is formulated as
\begin{equation}
\small
\mathcal{L}_{q, k^+, \{k^-\}} = -\log \frac{\exp(q{\cdot}k^+ / \tau)}{\exp(q{\cdot}k^+ / \tau) + {\displaystyle\sum_{k^-}}\exp(q{\cdot}k^-  / \tau)},
\label{eq:infonce}
\end{equation}
where $\tau$ is a temperature hyper-parameter (we set $\tau$ = 0.07 following the setting of \cite{he2020momentum}). From Eq. (\ref{eq:infonce}) we can see that, a smaller InfoNCE loss requires a smaller distance between $q$ and $k^+$ and larger distances between $q$ and $\{k^-\}$. This meets our demand of grouping similar samples and discriminate dissimilar ones.

As we can see in Fig. \ref{fig:moco}, our contrastive learning mechanism contains two encoders used for the query and the key, respectively. The two encoders share the same structure in our implementation and in experiments, we tested various popular networks. The details will be presented in Section \ref{sec::detail}. To train the two encoders, we use the momentum update method proposed in \cite{he2020momentum}, where the parameters of the key encoder, $\theta_\textrm{k}$, is calculated based on the parameters of the query encoder, $\theta_\textrm{q}$, as $\theta_\textrm{k} \leftarrow m \theta_\textrm{k} + (1 - m) \theta_\textrm{q}$, 

where $m$ is a momentum coefficient ($m$ = 0.999 in our experiment following the setting of \cite{he2020momentum}). The main idea behind momentum update is that using a queue as a dictionary makes the key encoder unable to update by back-propagation, and only $\theta_\textrm{q}$ is updated by back-propagation. Also, the momentum update makes the key encoder changes very slowly between different batches, and therefore the keys generated in different batches will not change sharply.

\subsection{Fingerprint Generation and Matching}
After training, the query encoder is used as the fingerprinter to extract fingerprints from audio (the key/momentum encoder is used only in training). Given an audio track, its fingerprints are computed as follows. First, the audio is divided into segments of length $T$ (equals to the audio length $T$ used in Section \ref{sec:contrast_learning}) with overlap. Each segment is then converted into a Mel-spectrogram and sent to the encoder to generate an encoded embedding. We call each embedding a sub-fingerprint, and all the sub-fingerprints of the input audio form the fingerprint of the audio.

Audio identification is finally accomplished by matching the fingerprint of query audio against fingerprints extracted from a reference audio set and stored in a reference fingerprint database. This is performed using a straightforward method, where for each sub-fingerprint of the query audio, we search linearly in the reference database to find its nearest neighbor measured by cosine distance. The reference audio, which has the most number of nearest neighbors with the query audio, is returned as the identification result. Please note that the focus of our current work is to study the potential of contrastive learning for AFP, and therefore we do not consider any sophisticated indexing techniques in fingerprint matching. We do not consider the case where the reference version of the query audio is not in the database. These will be left for future work.

\section{Experimental Settings and Results}

\subsection{Dataset}
\label{sec:dataset}
In this work, we consider the identification of both speech and music audio, and to this end, we collected two datasets to train and test our model. The first dataset we used is VoxCeleb2 \cite{Chung18b}, composed of short speech clips with the length ranging from 3.97s to 220.22s, extracted from interview videos uploaded to Youtube. VoxCeleb2 contains a development set VocCeleb2-Dev and a test set VoxCeleb2-Test, where VocCeleb2-Dev contains 1,092,009 utterances for 5,994 celebrities, and VoxCeleb2-Test contains 36,237 utterances for 118 celebrities. 
The second dataset used in our experiments consists of 345,000 in-house music tracks of various genres. These music tracks have a length of 10s to 900s and are divided into a training set Music-Train of 340,000 tracks and a test set Music-Test of 5,000 tracks.

In our experiments, we combined VoxCeleb2-Dev and Music-Train for the MoCo-based contrastive learning. This results in a final training set of 1,432,009 audio tracks and 4,527 hours in length in total. The testing of our model was performed on VoxCeleb2-Test and Music-Test separately. For VoxCeleb2-Test, we performed random degradation on each audio in the set and generated a query set of 36,237 degraded speech clips. These clips were compared against the original VoxCeleb2-Test for identification. For Music-Test, we first extracted a 10s clip from a random position of each audio in Music-Test and then performed random audio degradation on the clip. This results in a query set of 5,000 music clips of 10s, which were finally identified against the original test set to evaluate the performance.

\begin{table}[H]
 \small
 \centering
 
 \begin{tabular}{ccc}
     \toprule
     Degradation & Params & Description \\
     \hline
     White noise & [0, 0.08] & Noise intensity\\
     Pitch & [-5, +5]  & Semitone\\
     Speed &  [0.8, 1.2] & Factor\\
     Tempo & [0.8, 1.2] & Factor\\
     Highpass & 2000 Hz & Cutoff frequency\\
     Lowpass &  300 Hz & Cutoff frequency\\
     Echo & (0.8, 0.88), (60, 0.4) & Delay (ms) and decay pairs\\
     EQ$^*$ & Haitsma \textit{et. al} \cite{haitsma2002highly} & Equalization\\
     MP3$^*$ & 32 Kbps  & MP3 compression\\
     
     \bottomrule
 \end{tabular}
 \caption{Audio degradation methods and their parameters. The Echo degradation contains the adding of two echos successively. EQ and MP3 compression are used only in the testing stage to evaluate the generalization of our model to unknown attacks.}
 \label{tab:augmentation}
\end{table}

\subsection{Audio Degradation}
As shown in Section \ref{sec:contrast_learning} and \ref{sec:dataset}, both the training and testing of our model require the use of audio degradation. Specially, in the training stage, audio degradation is used as data augmentation solution to generate positive samples and teach our model to learn how to group similar items. In the testing stage, audio degradation is used to distort the query audio to help test the robustness of our system.

In the experiments, we implemented several audio distortions which have been commonly used in the AFP literature. The details are given in Table \ref{tab:augmentation}. When audio degradation is needed, each type of degradation has a chance of 30\% to be selected and the parameter of each selected degradation is randomly determined within the range given in Table \ref{tab:augmentation} (the Params column). The selected distortions are combined to attack the audio. Different from existing works (e.g., \cite{haitsma2002highly} and \cite{baez2020samaf}) which generally evaluated the system against each distortion separately, we trained and tested our model using a mixture of random degradation, which we argue is a better simulation of the real-world environment.

\subsection{Implementation Details}
\label{sec::detail}
As described in Section \ref{sec:contrast_learning}, our contrastive learning model takes as input batches of audio tracks. These tracks were extracted from random positions of the audio in the training set, and they were converted to mono and resampled at 16 kHz before used for training. The audio snippets in $\{x^{query}\}$ and $\{x^{key}\}$ all have a length of $T$ = 2.5s. Mel-spectrograms were calculated from these snippets using a frame length of 1,024 samples with an overlap of 256 samples, and the dimension of feature was set to 128. Therefore, each Mel-spectrogram has a size of 128 $\times$ 200. In the testing stage, query and reference fingerprints were both obtained by first converting the audio to mono and 16 kHz, and then partitioning the audio into segments of $T$ = 2.5s with an overlap of 15\%. A Mel-spectrogram was calculated from each segment using the same settings above and went through the query encoder to obtain the corresponding sub-fingerprint.

As shown in Fig. \ref{fig:moco}, our model contains a query encoder and a key encoder, which share the same model structure. In experiment, different existing models, including VGG-11, VGG-16 \cite{simonyan2014very}, ResNet-18 and ResNet-50 \cite{he2016deep}, all with ImageNet pre-trained parameters, and VGG-16 without ImageNet pre-training were tested and compared. These models were followed by a global average pooling layer and two fully connected layers (number of units = 256 for both layers) to form the encoder. The output of our encoder is a 256-dimensional vector, which was normalized by L2 norm to generate the desired embedding. An example of our encoder is shown in Table \ref{tab:encoder}.

Finally, in the training of our model, we used SGD as the optimizer with decay and momentum set to 0.0001 and 0.9 respectively. The batch size was 256. The initial learning rate was 0.03, and cosine learning rate schedule was utilized.

\begin{table}[H]
\small
\centering

\begin{tabular}{ccc}
    \toprule
    Layer & Shape & Output size  \\ 
    \midrule
    Input & 128$\times T$ &     \\ 
    conv2d - maxpool2d & 3$\times$3, 64 & 64, 64, $T$/2    \\ 
    ... & & \\
    conv2d - maxpool2d& 3$\times$3, 512 & 512, 8, $T/$32    \\ 
    global avg pooling & & 512, 1, 1        \\
    fully connected & 512$\times$256 & 256 \\
    fully connected & 256$\times$256 & 256 \\
    \bottomrule
\end{tabular}
\caption{An example of our encoder. The model takes a Mel-spectromgram as input, and is composed of a VGG-11.
}
\label{tab:encoder}
\end{table}

\subsection{Experimental Results}

We tested different encoders (fingerprinters) for their hit rate of audio identification on VoxCeleb2-Test and Music-Test. The results are shown in Table \ref{tab:result}. For comparison, we also implemented an existing algorithm based on which the Shazam system is built.

From Table \ref{tab:result} we can see that, the choice of encoders has a large impact on the performance of AFP for the VoxCeleb2 dataset, and among all the models we tested, the VGG-16 pre-trained with ImageNet achieved the highest hit rate. Especially, the VGG-based encoders outperformed the ResNet-based methods for all the cases, which indicates that, in the problem of AFP, using deeper networks and learning semantically more meaningfully features do not always lead to better performance. For Music dataset, the performance difference between encoders are notably smaller.
One possible reason is that music is more discriminative than speech and easier to recognize; another reason may be that the size of Music-Test is small.

Please note that since our query audio used in both test sets has been distorted by a combination of serious distortions, the Shazam-based methods performed rather poor. This proves from another point of view the powerfulness of our contrastive learning method for AFP.

As shown in Table \ref{tab:augmentation}, two types of distortions, i.e., equalization and compression have not been used for training, and during testing, the hit rates of our AFP system under these two attacks are all 100\% for both test sets. This indicates the generalization of our model against unknown attacks.

Finally, in addition to robustness, compactness is another important property of AFP. For a typical music track of the length of 3 min, the fingerprint extracted by our model has a size of about 400 KB when stored using 32-bit floating numbers. Therefore, a fingerprint database consisting of 100 million music tracks will occupy a storage size of about 400 GB, which can be hosted by a typical server. Our model has great potential to be used for industry applications.

\begin{table}[t]
 \small
 \centering
 
 \begin{tabular}{ccc}
     \toprule
     & \textit{VoxCeleb2-Test} & \textit{Music-Test}\\
     \hline
     Shazam-based & 0.00$\%$ & 3.40$\%$ \\
     VGG-11 & 80.67$\%$ & 99.34$\%$ \\
     VGG-16 & 84.54$\%$  & 99.62$\%$ \\
     ResNet-18 & 72.63$\%$  &  99.36$\%$ \\
     ResNet-50 & 74.06$\%$ & 99.74$\%$ \\
     VGG-16 (no pretrain) & 74.16$\%$ & 99.80$\%$ \\
     \bottomrule
 \end{tabular}
 \caption{Hit rate of different algorithms on \textit{VoxCeleb2-Test} and \textit{Music-Test}.}
 \label{tab:result}
\end{table}

\section{Conclusion}
In this work, we present a contrastive unsupervised learning method for AFP, which has proven by a set of experiments to be robust to serious audio distortions including the challenging speed change and pitch shifting. Future works include 1) the test of our algorithm on larger dataset; and 2) the use of more sophisticated indexing and retrieval techniques.

\vfill\pagebreak

\bibliographystyle{IEEEbib}
\bibliography{strings,refs}

\end{document}